\documentclass[preprint,showpacs,preprintnumbers,amsmath,amssymb]{revtex4}

\usepackage{graphicx}
\usepackage{dcolumn}
\usepackage{bm}

\begin{document}

\preprint{APS preprint}

\title{Temperature and interaction  dependence of the     moment of inertia  of a rotating condensate boson gas}

\author{Ahmed S. Hassan}

\author{Azza M. El-Badry}
\author{Shemi S. M. Soliman}
\affiliation{Department of Physics, Faculty of Science,  Minia  University, El Minia, Egypt.} 
\email{ahmedhassan117@yahoo.com}

\date{\today}

\begin{abstract}
In this paper, a developed Hartree-Fock semiclassical approximation is used to calculate the temperature and interaction  dependence of the moment of inertia of a  rotating condensate  Boson gas.  A fully classical and quantum mechanical treatment for the  moment of inertia  are given in terms of the normalized temperature.
 We  found that the moment of inertia is considerably affected by the interaction. The present analysis shows that the superfluid effects in the moment of inertia of a condensate Boson gas can be observed at temperatures $T  > 0.25 T_0 $ and not dramatically smaller than $T_0$.
\end{abstract}


\maketitle

\section{Introduction}
\label{intr} 
One of the most remarkable characteristics of the Bose-Einstein condensate (BEC)  is its response to rotate with  superfluid nature \cite{Matthews,Madison,exp,Dalibard3}.
The superfluid nature of this system  is investigated using the  moment of inertia. For a macroscopic system, the moment of inertia is given by the rigid value
unless it exhibits superfluidity. A deviation of the moment of inertia  from the rigid value represents an important manifestion of superfluidity.
In this respect, Stringari \cite{stringari1} drew a parallel between the rotating BEC and the superfluid systems, and he pointed out that the rotational properties of a BEC provides a natural way to analyze the deviations  from a rigid motion due
to condensation. Several studies showed that the evidence of the superfluidity in a rotating BEC is the reduction  of the  moment of inertia  below the classical rigid-body value \cite{stringari1,Brosens,Schneider,Odelin,Zambelli,Zambelli1}.

Mainly,  the moment of inertia is calculated in terms of  the effective {\it in situ} radii and the normalized temperature.
The approach of Brosens et al. \cite{Brosens} of the moment of inertia
is based on the  {\it in situ} radial radius $\langle x^2 + y^2\rangle$.
Their analysis focused on the difference of the moment
of inertia of a totally classical Boltzmann gas in a trap
and   $\langle x^2 + y^2\rangle $  for a Bose gas (cf.
Eq. (15)). Therefore, they missed the true superfluid effects
that may only be analyzed by calculating the moment of
inertia from quantum mechanical response to rotations.
In contrast, Stringari's work \cite{stringari1} is based on linear
response theory. He obtained the different contributions
from the condensate and the thermal cloud to the response
coefficient both for an ideal and an interacting Bose gas. Schneider et al. \cite{Schneider} presented a calculation of the fully quantum
mechanical moment of inertia for a microscopic cloud (in the presence of vortices)
of non-interacting atoms in a cylindrically symmetrical
trap. 
However, an ideal BEC of (non-interacting bosons) is not a true superfluid,
because the Landau criterion for superfluidity is not obeyed \cite{Annett}. Superfluidity, the formation of the
vortex lattice, is  a direct effect of inter-particle interactions, that would not
occur in the ideal BEC case.

\par
In this work, having clarified that the moment of inertia can be derived in terms of the effective  {\it in situ} radii,   we  discuss how quantitative results can be obtained in the presence of interatomic  interactions. The temperature-dependent for the {\it in situ radii} is calculated within the mean field Hartree-Fock
approximation \cite{dal,Sinha,Sandoval}. This approach can be summarized as follow:  a conventional method of statistical quantum mechanics is used to calculate  the temperature dependency {\it in situ} radii. The parametrized formula for the  {\it in situ} radii are used in calculating the moment of inertia. The obtained results 
showed that the above mentioned  quantities have a special  temperature behavior \cite{zha}.
 
 \par 
The  paper is planned as follows: section two includes  the basic formalism for  calculating 
 the effective in {\it situ} radii. 
Interaction and temperature dependency of the moment of inertia are given in section three. Conclusion  is given in the last section.
\section{{\it In situ} radii of interacting Bose gas}
The ideal Bose-Einstein condensation phenomenon is most conveniently described
in the grand-canonical ensemble.
For an ideal Bose gas, the average number of particles, $n_i$,
in a single particle state  $|i\rangle$ with energy $\epsilon_i$ is given by the
familiar Bose-Einstein distribution,
\begin{equation}
	n_i = \frac{{\textsc z} e^{-\beta \epsilon_i}}{1 - {\textsc z} e^{-\beta \epsilon_i}} 
\end{equation}
where $\beta=1/(k_B T)$,  $ {\textsc z} = e^{\beta \mu}$ is the effective fugacity, and $\mu$ is the chemical potential, 
determined by the conservation of total number of
particles
\begin{equation}
	N = \sum_{i=0}^\infty	n_i =  \sum_{i=0}^\infty \sum_{j=1}^\infty {\textsc z}^j e^{-j\beta \epsilon_i}
	\label{eq61-1}
\end{equation}
The degeneracy factors are
avoided by accounting for degenerate states individually. 
Once ${\textsc z}$ has been determined, all thermodynamically relevant
quantities can be calculated from partial derivatives
of the grand potential $q$, the logarithm of the grand
canonical partition function, such as
the 
{\it in situ radii}, the condensate fraction, etc.

The   effective {\it in situ} radius of trapped ideal boson gas was obtained by considering  the statistical quantum mechanics arguments \cite{zha,bra}.  For a trapped  boson in spherically symmetric harmonic potential, $V(r) = m\omega^2 r^2 /2$, the effective {\it in situ} radius  of a single particle   state $|i\rangle$ is given by its expectation value in this stat,i.e.
\begin{equation}
	\langle r_i^2\rangle = \frac{\epsilon_i}{m \omega^2} 
	\label{eq61}   
\end{equation}
with  $\epsilon_i = \hbar \omega (i + \frac{3}{2})$ is the eigenvalue of the potential $V(\bf r)$.
The effective {\it in situ} radius of $N$  atoms is  found
by gathering Eqs.(\ref{eq61-1}) and (\ref{eq61})
 \begin{eqnarray}
  N \langle r_i^2 \rangle 		&=&     \frac{1}{m\omega^2}\ \sum_{i=0}^\infty  \epsilon_i \sum_{j=1}^\infty  {\textsc z}^j  e^{-j \beta \epsilon_i}\nonumber\\
  &=&  - \frac{1}{m\omega^2}  \frac{\partial}{\partial \beta}  \sum_{i=0}^\infty \sum_{j=1}^\infty  \frac{{\textsc z}^j}{j}     e^{-j \beta \epsilon_i}
	\label{eq111}  
\end{eqnarray}  
 and can be expressed in terms of 
 the thermodynamic potential $q$, 
\begin{eqnarray}
q &=& - \sum_{i=0}^\infty \ln (1 - {\textsc z} e^{-\beta \epsilon_i}) =  \sum_{i=0}^\infty \sum_{j=1}^\infty \frac{{\textsc z}^j}{j}  e^{-j\beta  \epsilon_i }
\label{eq4-1}
\end{eqnarray}
the relation $\ln(1-y)  = - \sum_{j=1}^\infty \frac{y^j}{j}$ is used here.
Thus the effective {\it in situ} radius is given by
\begin{equation}
	\langle r^2 \rangle	 = - \frac{1}{m\omega^2}  \frac{\partial q }{\partial \beta}
	\label{r1}
\end{equation}

For a cylindrically symmetric trap with $\omega_x = \omega_y = \omega_\bot$, the temperature dependence of the three effective {\it in situ} is the same as in a spherically symmetric
trap discussed above.  Assuming an axial trap frequency $\omega_z = \lambda\omega_\bot$.
\begin{eqnarray}
N {\langle x^2\rangle} &=& - \frac{1}{3} \lambda^{1/3}  \frac{1}{m\omega_x^2}  \frac{\partial q }{\partial \beta} \nonumber\\
N {\langle y^2\rangle} &=& - \frac{1}{3}  \lambda^{1/3}     \frac{1}{m\omega_y^2}  \frac{\partial q }{\partial \beta} \nonumber\\
N {\langle z^2\rangle} &=& - \frac{1}{3}  \lambda^{-2/3}  \frac{1}{m\omega_z^2}  \frac{\partial q }{\partial \beta}
\label{eq105}  
\end{eqnarray} 
where $\lambda$ is the trap deformation parameter.
Generalization to
 highly anisotropic  trap (which is mainly used for rotating condensate) is straightforward. 
 Assuming that the trap deformation parameters for highly anisotropic trap are given by
 $$\lambda_x = \frac{\omega_z}{ \omega_x},  \lambda_y = \frac{\omega_z}{ \omega_y}$$
the  three effective {\it in situ} radii
 are given by \cite{zha,ahm1},
\begin{eqnarray}
N {\langle x^2\rangle} &=& - \frac{1}{3} \big(\frac{\lambda_x^2}{\lambda_y} \big)^{1/3}  \frac{1}{m\omega_x^2}  \frac{\partial q }{\partial \beta} \nonumber\\
N {\langle y^2\rangle} &=& - \frac{1}{3} \big(\frac{\lambda_y^2}{\lambda_x} \big)^{1/3}    \frac{1}{m\omega_y^2}  \frac{\partial q }{\partial \beta} \nonumber\\
N {\langle z^2\rangle} &=& - \frac{1}{3} \big(\frac{1}{\lambda_x \lambda_y} \big)^{1/3} \frac{1}{m\omega_z^2}  \frac{\partial q }{\partial \beta}
\label{eq107}  
\end{eqnarray}   

However, once the thermodynamic potential $q$  has been determined,  the effective {\it in situ}  radius  can be calculated.
 our approach
is expected to provide correctly the interaction dependence of $q$ potential, apart from the critical behavior near
the BEC transition temperature where the mean-field approach is known to fail.

 Generally, the simplest way to include the interaction effect  is to use 
 the Hartree-Fock approximation. 
 Within this approximation, the  thermal component is treated as
a gas of non-interacting atoms moving in a self-consistently determined mean-field potential given by
\begin{equation}
	V_{eff}(x,y,z) = V_{trap}(x,y,z) + 2 g [n_{th} (x,y, z) + n_0(x,y, z)],
	\label{eq3-1} 
\end{equation}
where $g = \frac{4\pi\hbar^2 a }{m}$ is the interaction strength and 
\begin{equation}
V_{trap}(x,y,z) =   \frac{1}{2} m [\omega_x^2 x^2  + \omega_y^2  y^2  +  \omega_z^2 z^2],
\label{eq2}
\end{equation}
with $\{ \omega_x,\omega_y, \omega_z\}$ are the effective trapping frequencies.
 The
densities of  the thermal  and condensate component are given as a solution of the two coupled equations:	
	the thermal atoms satisfies   Schr\"odinger equation
\begin{equation}
\Big[ \frac{ p_x^2+p_y^2+p_z^2}{2m} +   V_{eff}(x,y, z) \Big] 
 \psi_i(x,y, z) = \epsilon_i \psi_i(x,y, z) 
\label{eq7}
\end{equation}
while  the condensate part satisfies the time independent Gross-Pitaevskii equation
\begin{equation}
\Big[ \frac{ p_x^2+p_y^2+p_z^2}{2m} +   V_{eff}(x,y, z) - g  n_0(x,y, z) \Big]   \phi(x,y, z) = \mu \phi(x,y, z),
\label{eq8}
\end{equation}
Equations (\ref{eq7}) and (\ref{eq8})  along with the constraint that the total number of atoms $N$ is fixed,
\begin{equation}
	N = \int n_{th}{(x,y, z)}  dxdydz + \int n_0 {(x,y, z)} dxdydz
\end{equation}
form a closed set of equations which should be solved self-consistently. 
Two further simplifications can be made as a consequence of the relative diluteness
of the thermal component compared to the condensate \cite{Campbell}:\\ (i) at very low temperature  the effect of thermal atoms on the condensate can be  neglected. Therefore, setting  $n_{th} (x,y, z) \approx 0$ in Eq.(\ref{eq8}) and applying the Thomas-Fermi (TF) approximation gives
the usual TF profile for the condensate

\begin{equation}
	n_0(x,y, z) = \frac{ \mu - V_{trap}(x,y, z)}{g} 
	\label{eq10}
\end{equation}
For all $\mu > V_{trap}(x,y, z)$ and $n_0(x,y, z)$ = 0 elsewhere.
Substituting from Eq.(\ref{eq2}) in Eq.(\ref{eq10}) leads to, 
\begin{equation}
	n_0(x,y, z) =  \frac{\mu}{g} \Big[ 1 -  \frac{x^2 }{R_x^2(\mu)} - \frac{ y^2}{R_y^2(\mu)} - \frac{z^2}{R_z^2(\mu)}\Big]  
	\label{eq11}
\end{equation}
where
$
	R_\alpha(\mu) = \sqrt{\frac{ 2\mu}{m   \omega_\alpha^2}}  
	$
is the Thomas-Fermi radius at which the condensate density
drops to zero along the $x,y$ or $z$ axis. The
result  in Eq.(\ref{eq11}) can be expressed in terms of the condensate number of atoms through the relation between  $\mu$  and $N_0$,

\begin{eqnarray}
	N_0 &=& \int n_0(x,y, z) dxdydz \nonumber\\
	 	&=&  \frac{8\pi}{15} \frac{\mu}{g} (R_xR_yR_z) = \frac{8\pi}{15} \frac{\mu}{g} {\bar R}^3
	\label{eq12-1}
\end{eqnarray}
$\bar R$ representing the geometric mean $(R_xR_yR_z)^{1/3}$.
Equation (\ref{eq12-1}) can be inverted to give ${\mu}$ in terms of $N_0$ such that,
\begin{eqnarray}
	\mu  &=& \frac{1}{2} \hbar \omega_g \Big( \frac{15 N_0 a}{a_{har}}  \big)^{2/5} 
	\label{eq13} 
\end{eqnarray}
where $a$ is the s-wave scattering length, $a_{har}=\sqrt{\hbar/m \omega_g}$ and $\omega_g  = (\omega_x\omega_y \omega_z)^{1/3}$.\\
(ii) further, within the same approximation,  the mean-field energy, $2gn_{th}(x,y,z)$ due to the thermal
component itself can be neglected, so that the effective potential experienced by the thermal
atoms is then given by
\begin{eqnarray}
	V_{eff}(x,y,z) &=& V_{trap}(x,y, z)  + 2 g  n_0(x,y,z), \nonumber\\
	       &=& | V_{trap}(x,y,z)  - \mu_0| + \mu_0
				\label{eq14}
\end{eqnarray}
where the mean-field chemical potential $\mu_0$ is given by  Eq.(\ref{eq13}) and  $T \to  0$ limit is
indicated.
 Eq.(\ref{eq14}) shows that  the condensate density is drastically altered from the ideal case, reflecting that
the shape of the confining potential  has a three-dimensional `Mexican-hat' shape \cite{bretin}. Moreover, $\mu_0$ is  the relevant energy scale parametrizing the effects of interactions,
up to the point in the trap where $ \mu_0 = V_{trap}(x,y, z)$.

Now, it is straightforward  to calculate the 
thermodynamic potential $q$ for the interacting Bose gas \cite{Sinha,hau}. However, for
 large number of particles in the system, Eq.(\ref{eq4-1}) provides a complicated sum over $i$. It is hard
to evaluate this sum analytically in a closed form. Another possible way to do this analysis, is to use  the semiclassical approximation 
	 in which the sum in  Eq.(\ref{eq4-1})  is converted into a phase space integral \cite{Sinha,hau},
\begin{eqnarray}
	q   &=& q_0  + \frac{1}{(2\pi \hbar)^3} \sum_{j=1}^\infty \frac{{\textsc z}^j}{j}  \int e^{-j\beta[\frac{ p_x^2+p_y^2+p_z^2}{2m}  +  V_{eff}(x,y,z)]} {dp_xdp_ydp_z dxdydz}   \nonumber\\
&=& q_0  + \frac{1}{\lambda_{th}^3}   \sum_{j=1}^\infty \frac{{\textsc z}^j }{j^{5/2}} \int  e^{- j\beta  V_{eff}(x,y,z) } { dxdydz} 
	\label{eq5-3} 
\end{eqnarray}
where   $ q_0=-\ln (1-  {\textsc z}) $ is the thermodynamic potential accounted for the atoms in the ground state and $\lambda_{th} = \sqrt{\frac{2 \pi \hbar^2}{m k_B T}}$ is the thermal de Broglie wavelength. The second term in Eq.(\ref{eq5-3}) provides the thermodynamic potential for the thermal atoms.

In order to calculate the above integral (\ref{eq5-3}), we followed  the Hadzibabic and co-worker \cite{tammuz,Campbell} approach's and consider the same approximation.
For relatively high temperature,   (compared with $\mu_0/k_B$) the
majority of thermal atoms lie outside the  condensate in the region where \underline{ $V_{eff}(x,y, z) > \mu_0$
and  $ V_{eff}(x,y,z) = V_{trap}(x,y,z)$.} Therefore, it is  reasonable to approximate  the full effective potential
as the bare trapping potential and consider only the region outside the condensate.  This
does not mean that the effect of interactions may be neglected  as the chemical potential
has a value that differs substantially from the ideal value. Therefore Eq.(\ref{eq5-3}) becomes
\begin{eqnarray}
 q  &=& q_0 + \frac{1}{\lambda_{th}^3}\sum_{j=1}^\infty \frac{ 1}{j^{5/2}} \int  e^{-j\beta [V_{trap}(x,y,z)  - \mu_0]}  dxdydz
	\label{eq15-0}
\end{eqnarray} 
Note that in deriving this equation we used ${\textsc z}^j = e^{j\beta \mu_0}$.
  Substituting the harmonic
form of $ V_{trap}(x,y, z)$ into Eq.(\ref{eq15-0}) gives
\begin{eqnarray}
q  &=& q_0 + \frac{1}{\lambda_{th}^3}\sum_{j=1}^\infty \frac{{1} }{j^{5/2}} \int  e^{-j\beta [\frac{1}{2} m  (\omega_x^2 x^2 + \omega_y^2 y^2+  \omega_z^2 z^2 )  - \mu_0]}
	\label{eq15}
\end{eqnarray} 
Introducing the thermal radius, which fixed the maximum value of the chemical potential compared to $k_B T$, 

\begin{equation}
R_\alpha'(T) = \sqrt{\frac{ 2}{ \beta m  \omega_\alpha^2}}
	\label{rt}
\end{equation}
 these radius
 is  equivalent to  the condensate Thomas-Fermi radius at  which the thermal density
drops to zero along $T \to 0$.
In terms of $R_\alpha'(T)$ Eq.(\ref{eq15}) becomes,
\begin{eqnarray}
q   &=& q_0  +   \frac{1}{\lambda_{th}^3}\sum_{j=1}^\infty \frac{1 }{j^{5/2}}   \int  e^{-j\big(  \frac{x^2}{{R_x'}^2} + \frac{y^2}{{R_y'}^2}  +  \frac{{z}^2}{{R_z'}^2} - \alpha_0\big)}  dxdy dz \nonumber\\
 &=& q_0  + 4\pi \frac{{R_x'R_y'} R_z'}{\lambda_{th}^3}  \sum_{j=1}^\infty \frac{1 }{j^{5/2}} \int_{\sqrt{\alpha_0}}^\infty R^2  e^{-j(R^2 - \alpha_0)}  dR
	\label{eq15-1}
\end{eqnarray} 
where the factor $4\pi$ is due to the integration over the angles and
 \begin{eqnarray}
 \alpha_0 &=&  {\mu_0 \beta},\ \
	R^2 =    \frac{x^2}{{R_x'}^2} + \frac{y^2}{{R_y'}^2} + \frac{{z}^2}{{R_z'}^2},
\end{eqnarray}
 it is sensible to introduce the variable $Q$, where
\begin{equation}
{Q^2} = {R^2} -  \alpha_0
\end{equation}
to rewrite (\ref{eq15-1}) as
\begin{eqnarray}
q   &=& q_0  + 4\pi \frac{{R_x'R_y'} R_z'}{\lambda_{th}^3}  \sum_{j=1}^\infty \frac{1 }{j^{5/2}}  \int_0^\infty Q^2 \Big(1+ \frac{\alpha_0}{Q^2}\Big)^{\frac{1}{2}}  e^{-j \frac{Q^2}{2}}  dQ \nonumber\\ 
&=&  q_0  + 4\pi \frac{{R_x'R_y'} R_z'}{\lambda_{th}^3} \sum_{j=1}^\infty \frac{1 }{j^{5/2}}   \int_0^\infty  (Q^2 + \frac{\alpha_0}{2})  e^{-j {Q^2}}  dQ \nonumber\\ 
\label{eq106} 
\end{eqnarray}
where the binomial expansion has been evaluated to first order in $\alpha_0$ .
Evaluating the Gaussian integral in (\ref{eq106})
 and used
$\zeta(s)=  \sum_{j=1}^\infty \frac{1}{j^s}$ 
 leads to
\begin{eqnarray}
q   &=& q_0   + 4\pi \frac{R_x'R_y' R_z'}{\lambda_{th}^3}  \sum_{j=1}^\infty \frac{1 }{j^{5/2}} \Big(  \frac{\sqrt{\pi}/4}{j^{3/2}} +  \frac{\sqrt{\pi}/4}{j^{1/2}}\ \alpha_0  \Big) \nonumber\\
&=& q_0  +  \Big(\frac{1}{\beta \hbar \omega_g}\Big)^3 \big[\zeta(4)  + {\beta \mu_0}    \zeta(3) \big]
	\label{eq16}
\end{eqnarray}
where $\omega_g = (\omega_x\omega_y\omega_z)^{1/3}$ for highly anisotropic trap. Expressions for other trap type (cylindrically or spherically) can be extracted from Eq.(16) by setting the trap frequencies.
Direct comparison between this results and  the $q$ potential for the ideal system  \cite{pat}, 
\begin{equation}
q^{id}  = q_0^{id} +  \Big(\frac{1}{\beta \hbar \omega_g}\Big)^3  g_4({\textsc z})  
\label{eqhf1}  
\end{equation}
shows that the first and the second terms in Eq.(\ref{eq16}) are in comparable  with the result of the ideal system at $T < T_0$ with $T_0 = \frac{\hbar \omega_g }{k_B} \big(\frac{N}{\zeta(3)}\big)^{1/3}$ is the BEC transition temperature for the non-interacting gas  and  $\zeta$ is the Riemann zeta function and  $g_\nu(\textsc z)  = \sum_{k=1}^\infty {{\textsc z}}^k/k^\nu$ is the usual Bose function. The last term in Eq.(\ref{eq16}) accounted well for the interaction effect. This  effect can be seen more clearly by 
using Stringari et al. \cite{dal,strin} interaction scaling parameter $\eta$. This parameter is determined  by the ratio between the chemical potential at $T=0$ value calculated in Thomas-Fermi approximation, $\mu_0$  and the transition temperature for the non-interacting particles in the same trap, i.e. 
$\eta = \frac{\mu_0}{K_B T_0} $ (the typical values for $\eta$ for most
experiments ranges from 0.30 to 0.40.).

Finally, we reach to the main results of our work. The interaction dependence for the \ {\it in situ} radius for the spherically symmetric trap can be obtained by
substituting from Eq.(\ref{eq16}), after setting $\omega_x = \omega_y = \omega_z = \omega $, into Eq.(\ref{eq111}), i.e. 
\begin{eqnarray}
	N \langle r_i^2 \rangle  &=& -  \frac{1}{m\omega^2}  \frac{\partial q }{\partial \beta} \nonumber\\
	     &=&  -  \frac{1}{m\omega^2}  \frac{\partial }{\partial \beta} \big[-\ln (1-  {\textsc z}) +  \Big(\frac{1}{\beta \hbar \omega}\Big)^3 [\zeta(4)  + {\beta \mu_0}    \zeta(3)] \big]\nonumber\\
			&=&  - \frac{1}{m\omega^2}   \big[ \frac{-1}{1- {\textsc z}}\frac{\partial {\textsc z}}{\partial \beta}  -  \Big(\frac{1 }{\hbar \omega}\Big)^3  [3\frac{\zeta(4) }{\beta^4} + 2  \mu_0   \frac{\zeta(3)}{\beta^3}] \big]\nonumber\\ 
			&=&  \frac{\mu_0}{m\omega^2} N_0(T) + 3 \Big(\frac{k_BT}{m \omega^2}\Big) [\zeta(4)  + \frac{2}{3}\frac{\mu_0}{k_BT} \zeta(3)] \Big(\frac{k_B T}{\hbar \omega}\Big)^3\nonumber\\ 
			&=& \frac{\mu_0}{m\omega^2} N_0(T) + 3 \Big(\frac{k_BT}{m \omega^2}\Big) \big[\frac{\zeta(4)}{g_3({\textsc z})}  + \frac{2}{3}\frac{\mu_0}{k_B T} \frac{\zeta(3)}{g_3({\textsc z})}\big]  (N - N_0(T))
	\label{r7}
\end{eqnarray}
where  $ q_0=-\ln (1-  {\textsc z}), {\textsc z} = e^{\beta \mu}$  and $N_0(T) = \frac{{\textsc z}}{1-{\textsc z}}$  are used here. 
 
The generalization of the above treatment to a trap  with three different frequencies,
($\omega_x, \omega_y$ and $\omega_z$ ) is straightforward. Substituting from Eq.(\ref{eq16}) into Eq.(\ref{eq107}) leads to,
\begin{equation}
N \langle x^2 \rangle = \big(\frac{\lambda_x^2}{\lambda_y} \big)^{1/3}   \Big(\frac{k_BT}{ 3m \omega_x^2}\Big) \big\{  \frac{\mu_0}{k_BT} N_0(T) +     \big[3\frac{\zeta(4)}{g_3({\textsc z})} +  {2} \frac{\mu_0}{k_B T} \frac{\zeta(3)}{g_3({\textsc z})}\big] (N - N_0(T)) \big\}
	\label{r8}
\end{equation}
and analogously for $\langle y^2\rangle$ and $\langle z^2 \rangle$.  
The first term of $\langle x^2\rangle$ in the curly brackets  give the contribution arising from the particles in the condensate, while the second one is the
contribution from the non condensed atoms. Both of them are scaled as $\frac{1}{\omega_x^2}$. Unlike the non-interacting system for which the contribution arising from the non-interacting particles in the condensate is scaled as $\frac{1}{\omega_x}$.

 Result in Eq.(\ref{r8}) is a complementary to the Stringari \cite{stringari1} result for non-interacting system. 
In fact, this result constitute the main result which enables us to immediately calculate the interaction and temperature dependence for the moment of inertia.

\section{  Moment of inertia } 
 Fast rotating condensate  is
expected to exhibit superfluid properties at critical rotation velocity $\Omega_c$ \cite{fet1}.
 For $\Omega < \Omega_c$, following Dalvofo et al. \cite{dal},  the moment of inertia $\Theta$, relative to the $z$-axis, can be defined as the linear response of
the system to a rotational field $H_{ext} = -\Omega L_z$, according to the formula

\begin{equation}
\langle  L_z \rangle = \Omega \Theta
\end{equation}
where 
 the average here is taken on the state
perturbed by $H_{ext}$. For a rigid body rotation, the moment of inertia takes the  value

\begin{eqnarray}
	\Theta_{rig}&=&  m  N \langle x^2 + y^2\rangle \nonumber\\
	&=& m [\langle y^2 + x^2 \rangle_0 N_0 (T) +  \langle y^2 + x^2 \rangle_{nc} (N - N_0(T))]
	\label{ncon}
\end{eqnarray}
where
\begin{eqnarray}
	\langle y^2 + x^2 \rangle_0 &=&  \Big(\frac{k_BT}{3 m }\Big) \eta \big(1 - {\cal T}^3 \big)^\frac{2}{5} {\cal T}^{-1} \ [\big(\frac{\lambda_x^2}{\lambda_y} \big)^{1/3} \frac{1}{\omega_x^2}  + \big(\frac{\lambda_y^2}{\lambda_x} \big)^{1/3} \frac{1}{\omega_y^2}]\nonumber\\
	\langle y^2 + x^2 \rangle_{nc} &=&   \Big(\frac{k_BT}{3 m }\Big)   \{ 3 \frac{\zeta(4)}{\zeta(3)} + 2 \eta \big(1 - {\cal T}^3 \big)^\frac{2}{5}   {\cal T}^{-1} \} [\big(\frac{\lambda_x^2}{\lambda_y} \big)^{1/3} \frac{1}{\omega_x^2}  + \big(\frac{\lambda_y^2}{\lambda_x} \big)^{1/3} \frac{1}{\omega_y^2}]
	\label{r9}
\end{eqnarray}
 where the relation   
	$
	\frac{\mu_0}{k_B T} = \eta \big(1 - {\cal T}^3 \big)^\frac{2}{5} {\cal T}^{-1}
	$
	is used here, with ${\cal T} = \frac{T}{T_0}$ is the normalized temperature  
and $\eta$ is the Stringari interaction scaling parameter \cite{dal,strin}. 
Note that in (\ref{r9}) the appearance of $\omega_g/ \omega_x$ and $\omega_g/ \omega_y$
is due to the trap deformation effect. 
At low temperature,  the presence of a condensate pushes the thermal non-condensed cloud out, consequently increasing the effective {in situ} size of the
thermal component. 
At high temperatures $({\cal T} > 1 )$, the effect of the repulsive interaction becomes negligible as the density
of the Bose gas decreases dramatically with increasing temperatures.

While for $\Omega > \Omega_c$, the moment of inertia of the condensate is determined from the  quantum-mechanical arguments, in this case  $\Theta$ is given by,

\begin{equation}
 \Theta = \frac{2}{\cal Z} \sum_{i,j} \frac{|\langle j|L_z|i \rangle|^2}{E_i - E_j} e^{-\beta E_j}
\label{theta}
\end{equation}
where $\langle j|$ and $|i \rangle$ are eigenstates of the unperturbed
Hamiltonian, $E_j$ and $E_i$ are the corresponding eigenvalues and ${\cal Z}$ is the partition function.
The   Hamiltonian describing the interacting atomic gas in the potential (\ref{eq3-1})
is given by\cite {cooper}

\begin{equation} 
H = \frac{ p_x^2+p_y^2+p_z^2}{2m} +  V_{eff}(x,y,z) - \Omega  L,
\label{eq3}
\end{equation}
where $L_z$ is the angular momentum $L_z = xp_y - yp_x$. 
 the moment of inertia is explicitly evaluated by solving the equation
$$[H,X] = L_z$$
for the operator $X$, which according to (\ref{theta}), determines the moment of inertia through the
relation $ \Theta = < [L_z,X] >$. The explicit form of the operator $X$ is found to be \cite{Lipparin}

\begin{equation}
X = -\frac{i}{\hbar (\omega^2_x - \omega^2_y)} \sum_i [ (\omega^2_x - \omega^2_y) x_iy_i + \frac{2}{m} p^x_i p^y_i ]
\end{equation}
using the identity $<p_x^2> = m^2 \omega_x^2 < x^2>$ and $<p_y^2> = m^2 \omega_y^2 < y^2>$,  the moment of inertia takes the form

\begin{eqnarray}
	\Theta &=& \frac{m N }{(\omega^2_x - \omega^2_y)} {\big[ (\langle y^2 \rangle- \langle x^2 \rangle) {(\omega^2_x + \omega^2_y)} - 2 {(\omega_y^2 \langle y^2 \rangle- \omega_x^2 \langle x^2 \rangle}) \big]} \nonumber\\
	&=& m [  \epsilon_0^2  \langle y^2 + x^2 \rangle_0 N_0 (T) +  \langle y^2 + x^2 \rangle_{nc} (N - N_0(T))]
	\label{theta1}
\end{eqnarray}
Where
\begin{eqnarray}
\epsilon_0  &=& \frac{\langle x^2 - y^2 \rangle_0}{\langle x^2 + y^2 \rangle_0} \equiv \frac{\omega_y - \omega_x}{\omega_y + \omega_x}
\end{eqnarray}
 the indices $\langle   \rangle_0$ and $\langle  \rangle_{nc}$ in Eq.(\ref{theta1}) mean the average taken
over the densities of the Bose condensed and noncondensed
components {\it in situ}, respectively. The
quantity $\epsilon_0 $ is the deformation parameter of the condensate.

The above results for the moment of inertia, Eq.(\ref{ncon}) and Eq.(\ref{theta1}) were 
derived at non-zero temperature and for interacting system. Therefore, it is important
to investigate the dependence  of the moment of inertia  on these two  parameters. This dependency can be achived by considering
 the deviation
of the moment of inertia from its rigid-body value, i.e.
\begin{eqnarray}
	\frac{\Theta}{\Theta_{rig}} &=& \frac{\epsilon_0^2  \langle y^2 + x^2 \rangle_0 N_0 ({\cal T}) +  \langle y^2 + x^2 \rangle_{nc} (N - N_0(T)) }{ \langle y^2 + x^2 \rangle_0  {N_0 ({\cal T})} +   \langle y^2 + x^2 \rangle_{nc} (N - N_0(T))  }  \nonumber\\
	&=& \frac{\epsilon_0^2\ \eta \big(1 - {\cal T}^3 \big)^\frac{2}{5} {\cal T}^{-1} N_0 ({\cal T}) + N \{ 3 \frac{\zeta(4)}{\zeta(3)} + 2 \eta \big(1 - {\cal T}^3 \big)^\frac{2}{5}   {\cal T}^{-1} \} {\cal T}^3 }{ \eta \big(1 - {\cal T}^3 \big)^\frac{2}{5} {\cal T}^{-1}  {N_0 ({\cal T})} +  N \{ 3 \frac{\zeta(4)}{\zeta(3)} + 2 \eta \big(1 - {\cal T}^3 \big)^\frac{2}{5}   {\cal T}^{-1} \} {\cal T}^3  }\nonumber\\
	&=& \frac{\epsilon_0^2 \ \eta \big(1 - {\cal T}^3 \big)^\frac{2}{5}  [1 - {\cal T}^3]   + 2 \eta \big(1 - {\cal T}^3 \big)^\frac{2}{5} {\cal T}^{3} + 3 \frac{\zeta(4)}{\zeta(3)} {\cal T}^4 }{ \ \eta \big(1 - {\cal T}^3 \big)^\frac{2}{5}  [1 - {\cal T}^3]   + 2 \eta \big(1 - {\cal T}^3 \big)^\frac{2}{5} {\cal T}^{3} + 3 \frac{\zeta(4)}{\zeta(3)} {\cal T}^4 }
	\label{eq54}
\end{eqnarray} 
Result (\ref{eq54}) explicitly shows that at temperature greater than BEC transition temperature,
 where $N_0/N =0$, 
${\Theta}={\Theta_{rig}}$.
 While at $T=0$, it reduces to ${\Theta}= \epsilon_0^2  {\Theta_{rig}}$.
 
\begin{figure}
\resizebox{0.50\textwidth}{!}{\includegraphics{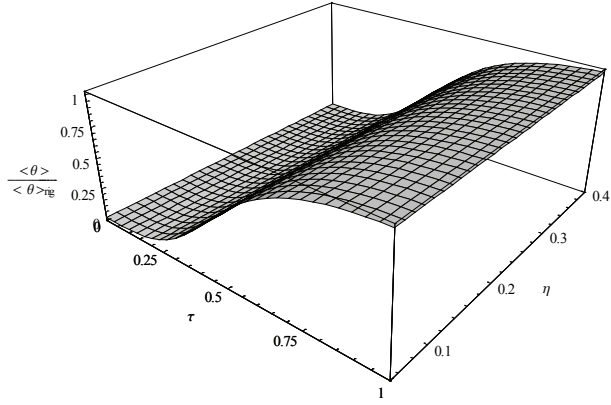}}
\caption{Moment of inertia ${\Theta}$ divided by its rigid value ${\Theta_{rig}}$, as
a function of ${\cal T}$ and $\eta$ for $\epsilon_0= - 0.032$. The trap parameters of Ref. \cite{exp3} are used. }
\label{f1}
\end{figure}

\begin{figure}
\resizebox{0.50\textwidth}{!}{\includegraphics{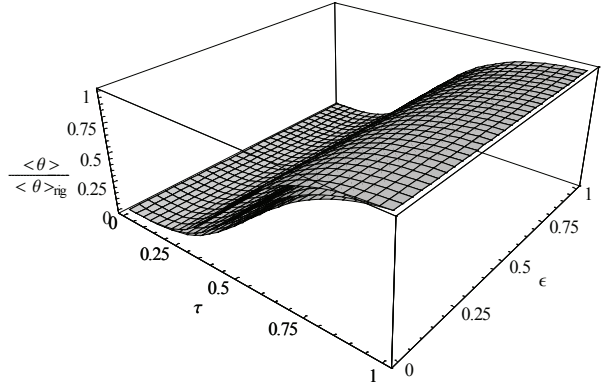}}
\caption{Moment of inertia ${\Theta}$ divided by its rigid value ${\Theta_{rig}}$, as
a function of the normalized temperature  ${\cal T}$ and the condensate deformation  $\epsilon_0$ for $\eta$ = 0.2, 0.4. and 0.6 from the bottom to top respectively. }
\label{f2}
\end{figure}

In Fig.(\ref{f1}), $\frac{\Theta}{\Theta_{rig}}$ is represented graphically as
a function of ${\cal T}$ and $\eta$ for $\epsilon_0=0.031$. The trap parameters of Ref. \cite{exp3} are used.
This figure shows that  $\frac{\Theta}{\Theta_{rig}}$   has a monotonically increasing nature due to the increase of the normalized temperature. 
This increase is minor in the small temperature range and is rapid in the intermediate temperature range. 

 Fig.(\ref{f2}) draws $\frac{\Theta}{\Theta_{rig}}$ as a function of $T$ and $\epsilon_0$ for different interaction parameter $\eta$.
This figure shows that, $\frac{\Theta}{\Theta_{rig}}$   has a monotonically increasing nature due to the increase of the normalized temperature. Moreover, the effect of interaction parameter is clear.

\begin{figure}
\resizebox{0.50\textwidth}{!}{\includegraphics{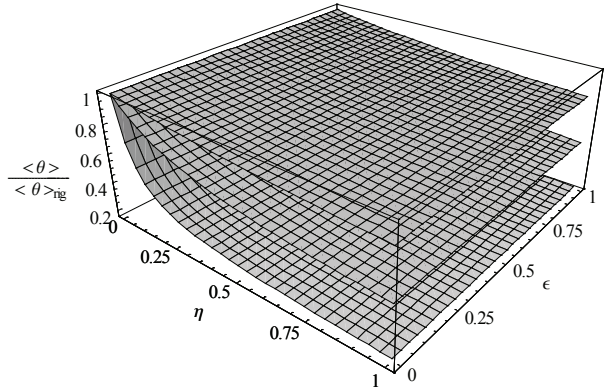}}
\caption{Moment of inertia ${\Theta}$ divided by its rigid value ${\Theta_{rig}}$, as
a function of the  condensate deformation parameter $\epsilon_0$, interaction parameter $\eta$ for normalized temperature $\tau =  0.4, 0.6$ and 0.8 from the bottom to top respectively. }
\label{f3}
\end{figure}

Fig(\ref{f3}) is devoted to illustrate  dependence of $\frac{\Theta}{\Theta_{rig}}$ on the interaction parameter $\eta$ and the condensate deformation parameter $\epsilon_0$ for different normalized temperature. This figure shows that the dependence of the  $\frac{\Theta}{\Theta_{rig}}$ on $\eta$ and  $\epsilon_0$
is considerably depended on the normalized temperature.

\section{Conclusion}
	In conclusion, we have shown that the moment of inertia  of a gas trapped by a harmonic potential can be explicitly
calculated in terms of the {\it in situ} radii and temperature dependence for both the condensate and thermal atoms.
Interaction  affected the value of the moment of inertia by changing its temperature dependence. An interesting feature is noticed that the moment of inertia ${\Theta}$ divided by its rigid value ${\Theta_{rig}}$ has a monotonically rapid increasing nature with the normalized temperature for $ 0.25 <\tau < 0.9$. The present analysis  recommended  Stingari  first important conclusion for non-interaction system which is the superfluid effects in the moment of inertia of a harmonically trapped Bose gas should be observable at temperatures
not dramatically smaller than the transition temperature for BEC.
 Our method can
be extended to investigate the moment of inertia for system of rotating  boson in a combined optical-magnetic trap. 

\end{document}